# Absence of dissipationless transport in clean 2D superconductors

A. Benyamini[1†*], E.J. Telford[2*], D.M. Kennes[3], D. Wang[2], A. Williams[2], K. Watanabe[4], T. Taniguchi[4], J. Hone[1], C.R. Dean[2], A.J. Millis[2], A.N. Pasupathy[2]

[1]Department of Mechanical Engineering, Columbia University, New York, NY, USA

[2]Department of Physics, Columbia University, New York, NY, USA

[3]Dahlem Center for Complex Quantum Systems and Fachbereich Physik, Freie Universitat Berlin, 14195 Berlin, Germany

[4]National Institute for Materials Science, 1-1 Namiki, Tsukuba, 305-0044 Japan

[*]Equally contributing authors

[†]Corresponding author: ab4442@columbia.edu

**Dissipationless charge transport is one of the defining properties of superconductors (SC). The interplay between dimensionality and disorder in determining the onset of dissipation in SCs remains an open theoretical and experimental problem. In this work, we present measurements of the dissipation phase diagrams of SCs in the two dimensional (2D) limit, layer by layer, down to a monolayer in the presence of temperature ($T$), magnetic field ($B$), and current ($I$) in 2H-NbSe$_2$. Our results show that the phase-diagram strongly depends on the SC thickness even in the 2D limit. At four layers we can define a finite region in the I-B phase diagram where dissipationless transport exists at $T = 0$. At even smaller thicknesses, this region shrinks in area. In a monolayer, we find that the region of dissipationless transport shrinks towards a single point, defined by $T = B = I = 0$. In applied field, we show that time-dependent-Ginzburg-Landau (TDGL) simulations that describe dissipation by vortex motion, qualitatively reproduce our experimental I-B phase diagram. Last, we show that by using non-local transport and TDGL calculations that we can engineer charge flow**

**and create phase boundaries between dissipative and dissipationless transport regions in a single sample, demonstrating control over non-equilibrium states of matter.**

A uniform superconducting condensate can transmit kinetic energy from one end to the other via electrical currents without any dissipation[1]. The dissipationless energy transfer is disrupted by the motion of vortices which can arise due to an applied magnetic field, thermal or quantum fluctuations, or electrical current[2–4]. Experimentally the magnetic field and electrical current also act as two control knobs to probe and alter the superconducting state[5]. Magnetic field tunes the vortex density, $n_V = \frac{B}{\Phi_0}$ ($\Phi_0 = \frac{h}{2e}$ quantum of flux attached to a single vortex, $h$ planck's constant, $e$ electron charge) and electrical current exerts a Lorentz force on each vortex, $F_{Lorenz} = J\Phi_0 t$ ($t$ sample thickness and vortex length, $J$ current density and the force is perpendicular to the magnetic field and the current). In the classical 2D mean-field picture at $T = 0$ and in the absence of pinning, vortices will form a solid due to vortex-vortex interactions. In the presence of a current the Lorenz force will move the solid, causing energy dissipation[2], once the force overcomes the vortex solid confining potential at the sample boundaries. Defects inevitably present in real materials may pin vortices, so that no dissipation would be measured until the Lorenz force overcomes the pinning potential. At low vortex density vortices will be pinned individually while at high densities, where vortex-vortex interactions dominate, vortices will be pinned collectively in a solid or glass like states[4,6,7]. Beyond mean-field physics, thermal and perhaps quantal fluctuations can melt the solid and a non-zero current can dislodge vortices from the pinned solid/glass, and create vortex-antivortex pairs, leading to current-dependent dissipation at zero and nonzero applied field. Above the melting temperature pinned vortices will be thermally activated, contributing to dissipation[4,6]. Figure 1b summarizes the expected B-T phase diagram in 2D in the linear-regime in the presence of pinning[6]. Understanding how vortices, the topological defects of a superconductor, are created and proliferate in a flowing current beyond the linear regime or even the existence of a linear regime in 2D are important problems in fundamental physics. At the same time, dissipation in 2D SCs is crucial to applications that depend on zero resistance and phase coherence in superconductors.

The many-body vortex state is complex due to the interplay of long ranged vortex-vortex interactions, pinning, thermal and quantum fluctuations. In 2D, one observes interesting phenomena such as the SC-insulator transition[8,9] and metallic-like behavior in the SC state[10]. Experimentally, in thin film SCs produced by evaporation or sputtering, crystal imperfections are observed to increase with reduced thickness[11–13]. Ultra-thin 2D superconductors produced in this way are therefore typically in a disorder-dominated limit. Van der Waals materials are 2D in nature and can be exfoliated to a single layer with the same level of crystal imperfection as in bulk[14]. These material offer both a unique opportunity to study ultra-thin SCs beyond the disorder/pinning dominated limit and new perspectives into dissipation mechanisms in 2D SCs.

The equilibrium B-T phase diagrams of thin crystalline SCs (the limit of small current drive) has been measured in a variety of van der Waals SCs[10,14]. It was recently demonstrated that at high values of current drive and moderate magnetic fields, a dissipative state with nonzero resistance emerges as $T \to 0$[15]. A natural question arises: at what magnetic fields and currents does one make a transition from a dissipative to a dissipationless state at T=0? Here, we answer this question by investigating transport in 2H-NbSe$_2$ at thickness ranging from one to four layers, and complement the measurements with theoretical and numerical analyses based on TDGL.

The devices are fabricated by the dry transfer[16] and via contact technique[17] where an insulating hBN with embedded metallic contacts is used to pick-up a few layer 2H-NbSe$_2$ and then placed on a second hBN, all done in an inert nitrogen environment in a glovebox. This allows one to simultaneously contact the air-sensitive 2H-NbSe$_2$, while preserving it from oxidation. As a result our samples are in the low disorder limit, with a mean free path $l > \xi_\parallel$ ($l = \frac{h}{e^2} \cdot \frac{1}{\sqrt{2\pi n_1 R_1}}$ is the 2D mean free path, $R_1$ and $n_1$ are the resistance and electron density per layer, and $\xi_\parallel$ is the in-plane Ginzburg-Landau coherence length). The sample parameters are summarized in S1 for the main devices in the paper. Illustrations of the device geometry are shown in figure 1a. All samples are in the 2D limit with a thickness smaller than the c-axis SC coherence length, $t < \xi_\perp \sim 2.7$nm[18]. We measure 4-probe voltages in a Hall bar-like geometry. We source and drain alternating current (AC) and direct current (DC) far from the measurement probes to measure the response to spatially uniform currents. We denote $R_{AC} = \frac{dV}{dI}\big|_{I_{DC}}$ and $R_{DC} = \frac{V_{DC}}{I_{DC}}$ for the inferred resistances. In supplementary S2 we show that a simple heating picture due to the finite DC currents in our measurements does not explain our results.

We begin our discussion by presenting data obtained in the limit of very weak applied currents. The measurements are limited by the experimental noise floor and in most regimes reasonably extrapolate to the linear response limit of current tending to zero for the AC excitation used in the experiments. Figure 1c shows the temperature dependence of the linear response resistivity at $B = 0$. The traces show a temperature above which the resistivity takes the normal state value, which we identify as the mean field transition temperature $T_c$ and a temperature at which the resistivity becomes indistinguishable from zero, which we identify as the Berezinskii-Kosterlitz-Thouless (BKT) transition temperature $T_{BKT} = \frac{\pi}{2}\rho_S(T = T_{BKT})$ where $\rho_S(T)$ is the temperature dependent superfluid stiffness of the 2D system[19]. The difference between $T_{BKT}$ and $T_c$ is a measure of the strength of beyond-mean-field fluctuations which we express in terms of $\eta$, proportional to the ratio of $T_c$ and the zero temperature superfluid stiffness: $\eta = \frac{2T_c}{\pi \rho_{S0}}$. In bulk 2D materials the superfluid stiffness is much larger than the transition temperature (fluctuation parameter $\eta \ll 1$). Using the standard mean field temperature dependence $\rho_S(T) = \rho_{S0}\left(1 - \frac{T}{T_c}\right)$ and the data in Fig. 1b yields $\eta \approx 0.11, 0.32, 0.66$ for the quadrilayer, bilayer and monolayer devices respectively. Use of the transition temperatures given in the supplemental material then gives for the superfluid stiffness per layer $\rho_{s0,layer} \approx 8.9, 5, 3.3 K$ for the quadrilayer, bilayer and monolayer devices respectively. The large values found here for the fluctuation parameter and small values of the superfluid stiffness per layer reflect the unique fragility of the superconducting state, particularly for the mono and bilayer systems.

We now turn to the full field and temperature dependence of the linear response resistivity, shown as a color map for the quadrilayer sample in Fig 1d. Consistent with the relatively small value of the fluctuation parameter, the results are consistent with mean field theory and with previous measurements on bulk crystals[20]. At each field, a reasonably sharp crossover separates a normal state with a resistivity that is very weakly dependent on field and temperature from a state with a resistance which is very low, and strongly field and temperature dependent, which we identify as the SC state. The crossover defines the upper critical field $H_{c2}(T)$ (defined here as the field at which the resistivity is 90% of the normal state value) shown by the white dashed line. Monolayer and bilayer devices display similar behavior, but with lower transition temperatures and correspondingly lower critical fields, and much broader crossover regimes as seen in figure 1e for the monolayer device.

Below $H_{c2}(T)$ the resistance is thermally activated down to our noise floor (figure S2a). The activation energies, $U$, are found (S3) to vary with the magnetic field as $U(B) = U_0 \cdot \log\left(\frac{B_0}{B}\right)$ as expected from the logarithmic vortex-vortex interactions. The prefactor $U_0$ is expected to arise from vortex-vortex interactions primarily mediated by the superfluid stiffness; the theoretical result is $U_0/k_B = \pi\rho_S$. Results of fitting measured resistivities are shown in the inset of Fig 1d and are consistent with the results found from the analysis of the $B = 0$ resistivity and further confirm the small values of the superfluid stiffness and the approximate linearity in layer number.

We now turn to the current dependence of the dissipation. Figure 2a shows for a monolayer and quadrilayer device the $I_{DC}$ dependence of the differential resistance $R_{AC}$ in log-scale obtained at $B = 0$ and the lowest temperature in this study, $250mK$. The $R_{AC}$ vs $I_{DC}$ curves are independent of temperature below $\sim 1K$ and we take the result as representative of the $T = 0$ behavior. The differential resistance curves indicate two characteristic drive currents. The lower drive current, $I_c$, is the drive at which the measured differential resistance becomes larger than the noise floor. The larger drive current, $I_0$, is the drive at which the differential resistance goes over to the normal state value. We interpret $I_0$ as the `microscopic' critical current marking the destruction of superconductivity. There are two physical origins of $I_0$: the `depairing current' for which the current excites quasiparticles over the gap, and the Ginzburg-Landau critical current related to current-induced gradients of the superconducting phase. In strongly type II materials such as the ones studied here the Ginzburg-Landau critical current is typically lower and controls the behavior. In the clean, low-T limit the Ginzburg-Landau critical current per layer is proportional to the transition temperature and to the square root of the superfluid stiffness per layer. $I_c$ is the current needed either to create a measurable number of vortex-antivortex pairs out of the condensate or to detach a measurable number of vortices from the pinned vortex lattice. Both phenomena are controlled by the superfluid stiffness of the device. Consistent with the relatively small value of the fluctuation parameter, the $B = 0$ $R_{AC}$ vs $I_{DC}$ trace shown in figure 2a for the quadrilayer device is as expected by mean field behavior: $I_c$ and $I_0$ coincide, indicating a discontinuous change of differential resistance from the noise floor to the normal state value. In contrast, the monolayer device shows a continuous onset of resistance from the noise floor consistent with the much lower superfluid stiffness. A summary of the critical currents with layer number is shown in terms of their densities in S4.

We now consider the magnetic field dependence of the dissipation. A peak in the differential resistance typically occurs at the Ginzburg-Landau critical current; this appears as a red region in figures 2b-2d. We see that $I_0$ decreases with $B$, initially linearly and then with some curvature. Smooth evolution of the $B = 0$ behavior is observed to finite magnetic field for all samples. Traces of the magnetic field dependence of $R_{AC}$ at $I_{DC} = 0$ are shown in the insets. Similarly to the DC current dependence of the resistance it is again observed that as the layer thickness is decreased the on-set of resistance occurs at a lower magnetic field, $B_c$ (noted by a white arrow in the insets), as $B_c/H_{c2} \sim 0.7, 0.2, 0.03$ for the quadrilayer, bilayer and monolayer respectively. This is also consistent with the much lower superfluid stiffness for lower layer number, as the on-set of resistance has to do with shaking vortices loose from the vortex lattice that is held by a force proportional to the superfluid stiffness. For the quadrilayer, at $B < H_{c2}$, the resistance versus current is characterized by a sharp onset from the noise floor near the critical current $I_0$. It is reasonable to assume that this sharp drop is to a dissipationless state, indicating that a substantial region of the I-B map corresponds to a dissipationless SC. For the bi and monolayer we observe that the noise-floor region in the I-B map shrinks quickly. In the monolayer, no sharp drops are seen in the resistance versus current down to the noise floor at $\sim 0.1 \cdot I_0$ at $B = 0$. This result, that dissipationless transport in clean monolayer 2H-NbSe2 with low superfluid stiffness exists only in the limit of $I = B = T = 0$ is the main finding of this paper.

For all samples, we observe $I_c$ to smoothly evolve from large applied fields to $B = 0$. In the limit of $B = 0$, no field-induced vortices exist in the system. At $B = 0$ the dissipation arises from the creation of vortex-antivortex pairs out of the condensate[21]. The creation process relies on the fact that a current pushes vortices and antivortices apart in opposite directions with a force that is independent of vortex-vortex separation; this force competes with the vortex-vortex attraction $\propto \frac{\rho_s}{r}$. The forces are equal at a distance $r \propto \frac{\rho_s}{J}$ defining an energy barrier $\propto \rho_s ln \frac{\rho_s}{J}$; over which the vortex-antivortex pairs must be activated. The rapid decrease of superfluid stiffness with layer number then shows why the dissipation effects are much more evident in the monolayer.

The dissipative state at intermediate currents, below $I_0$, is accompanied by a saturation of the resistance at the limit of $T = 0$ indicative of metallic-like behavior, figure 3a. At nonzero magnetic fields, we can gain physical intuition about the nature of the non-equilibrium metallic-like state, regime of intermediate resistance in figure 3b, using TDGL simulations (see S5-S6). Figure 3d shows the full non-equilibrium phase-diagram for a theoretical device geometry very similar to the experimental one; $40\xi \times 40\xi$ simulated compared to $\sim 120\xi \times 120\xi$ in the experiment. It exhibits qualitatively similar features to the experimental phase-diagram of a quadrilayer, figure 3b with a fluctuation regime which is negligible at $B = 0$ and broadens rapidly as field increases. The 90% line defining the `microscopic' $H_{c2}$ is very similar between experiment and theory. We see that at higher fields, a regime of nonvanishing dissipation is observed even at low drive currents. The non-vanishing dissipation comes from vortices which are detached from the vortex lattice and can move freely. This increased dissipation at lower current and higher field is more similar to the monolayer phase-diagram indicating that it is easier to detach vortices from the lattice at the monolayer limit.

We now add disorder (see S7). Figure 3c show the two simulated traces of resistance versus inverse temperature which show the same metallic-like behavior observed experimentally, figure 3a. The agreement to experiment solidifies the vortex dissipation picture at high field, which consists of a mixed vortex state of thermally activated pinned vortices and unpinned freely moving vortices[6]. In the simulations we observe the two vortex states, as well as how the freely moving vortices interchange with pinned vortices even without thermal fluctuations due to a strong enough Lorentz force, see supplemental movies M4-M11.

The activated region in the temperature-dependent resistance at high field can also be used to gain new insight into the depinning mechanisms at play in 2D SCs from the dependence of the activation energy on current and field. Many theoretical works have studied depinning in 2D, predicting a power-law dependence in the weak collective pinning[6,22]. To the best of our knowledge these theoretical predictions have never been explored experimentally in the clean 2D limit. Shown in Figure 4a is the measured activation energy dependence on $I_{DC}$ in a log-log scale for several magnetic fields for the quadrilayer device (for individual traces of resistance versus temperature, see supplementary information S8). We clearly see two power-law regimes across samples, separated by a current we denote by $I_1$ (or current density $J_1$). The magnetic field dependence of the fitted exponents for both regimes are summarized in figure 4b showing a logarithmic dependence on $B$. Following the theories[6,22], if an increase in the exponent indicates an increase in the bundle size, we can deduce that at low drive currents it is favorable to activate single vortices for any field, while at higher currents we cross to a regime where it is favorable to activate bundles with a size that increase with magnetic field. We can collect our observations at finite drive current to construct an I-B phase diagram for dissipation in 2D NbSe$_2$ shown in figure 4c. In dark and light blue are the two activated regimes of the pinned vortex state (dark for weak dependence on current and light for strong), in yellow the metallic-like regime of the unpinned vortex state and in red the normal state. Our observations show that as the thickness is reduced to the monolayer limit, the pinned vortex state regime shrinks until it eventually disappears at the monolayer thickness and any finite current at finite fields will detach vortices from the lattice creating dissipation.

Looking at the non-equilibrium phase diagram in figure 4c we recognize that if we park at a nonzero magnetic field and vary the current density in space, regions of different vortex states will be established. Realizing this will demonstrate non-equilibrium control over quantum matter[23,24]. In figure 5 we demonstrate how we stabilize different non-equilibrium steady-states of the 2D SC along the sample by sourcing non-uniform currents. Panels a-d show the experimental non-local response for increasing DC source current. For low DC current all non-local probes show activated behavior. As the DC current is increased, the probes closest to the source-drain contacts show saturated behavior while the furthest still show activated behavior. At the highest DC current, the source-drain area is in the normal state while the other regions are saturated.

To gain intuition on the way non-uniform currents affect the SC and vortices non-locally, we simulated this scenario with TDGL in the absence of pinning, see supplemental movies M12-M15. Pictures from the movies are shown in figure 5e for a finite field and different currents, the source and drain are noted by brown rectangles, color represents the size of the SC gap with blue being zero, and the white arrow's direction and length indicate the vortex velocity direction and size correspondingly. Following figure 5a-d the panels in figure 5e were generated with increasing DC current from left to right. An area where SC is destroyed next to the contacts is observed which increases in size as the current is increased. As no pinning sites are present vortices move freely with an overall vortex velocity in a direction that is perpendicular to the local current density and proportional in size to the local Lorentz force, $\vec{v} = \frac{\vec{F}_{Lorentz}}{\eta}$ ($\eta$ is the vortex viscosity). The non-uniform current density makes vortices move faster where the current density is higher and slower where the current density is lower, noted by the white arrows size. In the presence of pinning, vortices will get pinned if the combined Lorentz force and the force from other vortices is smaller than the pinning force. To map the simulation with no pinning to the case with pinning we need to imagine that the slower vortices will get pinned and be thermally activated, located further from the source drain, while the faster ones, close to the source drain, will move freely.

To summarize, few layer crystalline 2H-NbSe$_2$ enables us to investigate the physics of clean-limit, ultra-small superfluid stiffness superconductors. We find that the critical current density decreases quickly with lower layer number, making the samples sensitive to perturbations[15] at finite and even zero magnetic field in the monolayer limit. It is still unclear how this sensitivity depends on specific material parameters and if this may explain metallic-like behavior in other materials. As new extremely clean 2D SCs are found, as magic angle bilayer graphene[25] for example, the extreme clean limit may be achieved, $l_{mfp} \gg \xi_\parallel$ ($l_{mfp}$, electron mean free path), where the quantum mechanics of vortices dominates and new physical regimes may emerge such as a quantum vortex liquid[26], a quantum Hall fluid of vortices[27,28] and a fractional quantum Hall-like states[29]. An intriguing avenue of future research concerns the question what would emerge in this limit at non-equilibrium or when strongly coupled to other states[23,24]. It is also of interest if the non-equilibrium control of interfaces between different vortex states may be used to dynamically steer novel emergent physics.

**Acknowledgements**

We deeply thank Dani Shahar, Daniel Rhodes and Valerii Vinokour for fruitful discussions and input. This research was primarily supported by the NSF MRSEC program through Columbia in the Center for Precision Assembly of Superstratic and Superatomic Solids (DMR-1420634), the Global Research Laboratory (GRL) Program (2016K1A1A2912707) funded by the Ministry of Science, ICT and Future Planning via the National Research Foundation of Korea (NRF), and Honda Research Institute USA Inc. We acknowledge computing resources from Columbia University's Shared Research Computing Facility project, which is supported by NIH Research Facility Improvement Grant 1G20RR030893-01, and associated funds from the New York State Empire State Development, Division of Science Technology and Innovation (NYSTAR) Contract C090171, both awarded April 15, 2010. AJM and DMK were supported by the Basic Energy Sciences Division of the U.S. Department of Energy under grant DE-SC0018218. DMK additionally acknowledges support by the Deutsche Forschungsgemeinschaft through the Emmy Noether program (KA 3360/2-1).

**Author information**

**Contributions:**

The experiment was designed by AB and EJT, devices fabricated by AB, EJT and AJW, data taken by AB, EJT and DW, analysis by AB and EJT, theory and simulation by DMK and AJM, hBN crystals grown by KW and TT. All authors contributed equally to the manuscript.

**Competing interests:**

The authors declare no competing interests.

**Corresponding author:**

Avishai Benyamini – ab4442@columbia.edu

**Captions:**

**Figure 1: Equilibrium phase diagram of a 2D superconductor. a.** Illustration of a fully encapsulated 2H-NbSe2 device including the measurement setup. **b.** Illustration of the equilibrium phase-diagram of a 2D superconductor with pinning. The normal state is shown in light red, the pinned vortex state in light blue and a vortex solid or glass state at $T \to 0$ in light green. **c.** Temperature traces of $R_{AC}$ normalized to the normal state resistance, $R_N$, at $B = 0$ for the quadrilayer, bilayer and monolayer devices with a reference to TDGL simulation showing the expected behavior from mean-field at linear response. **d.** 2D color map of $R_{AC}$ for the quadrilayer device as a function of temperature and magnetic field. Dashed line shows the $0.9 \cdot R_N$ line. Inset: Dependence of the vortex dislocation energy scale $U_0$ on layer number. A linear fit, $U_0 = \varepsilon_0 \cdot N$, crossing the origin gives $\varepsilon_0 = 13.44\ K$ per layer. From the theoretical form[30], $U_0 = \frac{c_m \Phi_0^2 t}{\mu_0 \lambda^2}$ ($c_m$ is a model dependent constant, $\mu_0$ the vacuum permeability and $\lambda$ the bulk penetration depth), we get $c_m \sim 5.6 \cdot 10^{-3}$ assuming $\lambda = 250nm$ for bulk 2H-NbSe2[31] and $t_0 = 0.62nm$ the thickness of a single layer[32], which is of the order of the vortex-vortex interaction form[4], $c_m = 6.3 \cdot 10^{-3}$, and an order of magnitude larger than of the dislocation mediated 2D melting[30], $c_m \sim 2.4 \cdot 10^{-4}$. **e.** 2D color map of $R_{AC}$ for the monolayer device showing much wider transitions with respect to the quadrilayer device.

**Figure 2: Absence of dissipation less transport in a 2D superconductor. a.** Comparison of the current induced differential resistance of monolayer and quadrilayer devices emphasizing the on-set of resistance. Black arrows donate the current at which differential resistance is observed above the noise floor, $I_c$. **b-d.** Full B-I colormaps of $R_{AC}/R_N$, shown in log-scale, for a monolayer, bilayer and quadrilayer devices. The two critical currents $I_c$ and $I_0$ are noted at $B = 0$.

**Figure 3: TDGL simulation reproduce metallic-like behavior and main non-equilibrium experimental features**. **a, c.** Metallic-like behavior in experiment (quadrilayer 024) and from TDGL simulation including 100 pinning sites. Resistance is shown in log-scale vs inverse temperature traces for two different currents and magnetic fields noted in the panels and on the phase diagrams, panels b and d, by corresponding light and dark gray dots. **b, d.** Corresponding non-equilibrium phase diagrams to panels a,c. Insets show zoom-ins on the low magnetic field and high current regime.

**Figure 4: Vortex size controlled by magnetic field and current. a.** Activation energy dependence on $I_{DC}$ is plotted for several magnetic fields. Black arrows point to the estimated cross-over current. Dashed lines show power-law fits to the higher current regime. **b.** Dependence of the exponent extracted from the power-law fits for both low (gray) and high current (black) regimes. Dash lines shows a fit to $\alpha(B) = \alpha_0 \cdot \ln\left(\frac{B}{B_\alpha}\right)$ with $\alpha_0 = 0.034$ and $B_\alpha = 2.88T$ in the low current regime and $\alpha_0 = 0.205$ and $B_\alpha = 3T$ for the high current regime. **c.** Illustration of the non-equilibrium phase-diagram summarizing the different phenomenological regimes observed and the inferred theoretical physical pictures.

**Figure 5: Non-equilibrium real-space control over the superconducting state. a-d.** Normalized $R_{ac}$ as a function of $1/T$ measured at different distance from the source-drain electrodes for four DC currents. Illustrations of the device use the color scheme used in figure 4d for the different non-equlibrium steady-states. Vortices are illustrated by black points with arrows indicating their direction of motion. Panel b for example, show that the region closer to the source drain path is in the unpinned vortex state while the furthest is in the pinned vortex state. **e.** Frames from simulated vortex dynamics movies, see supplementary materials, for a non-uniform current at a fixed magnetic field for four different current regimes.

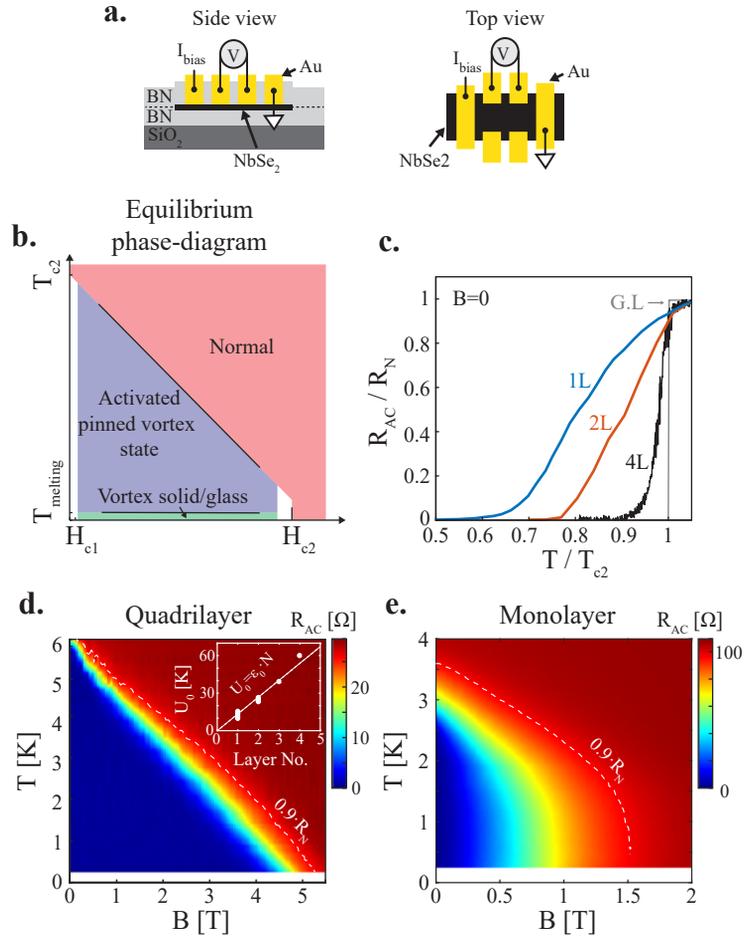

Figure 1

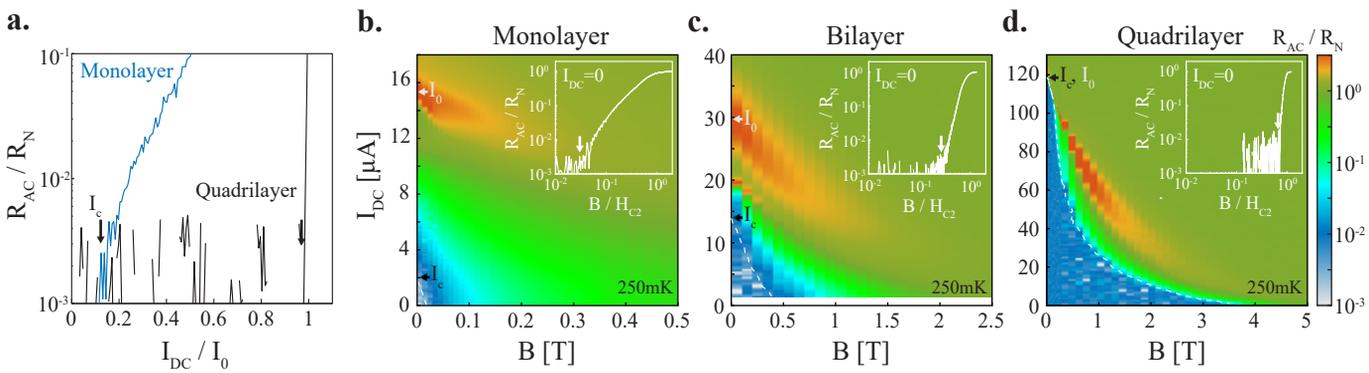

Figure 2

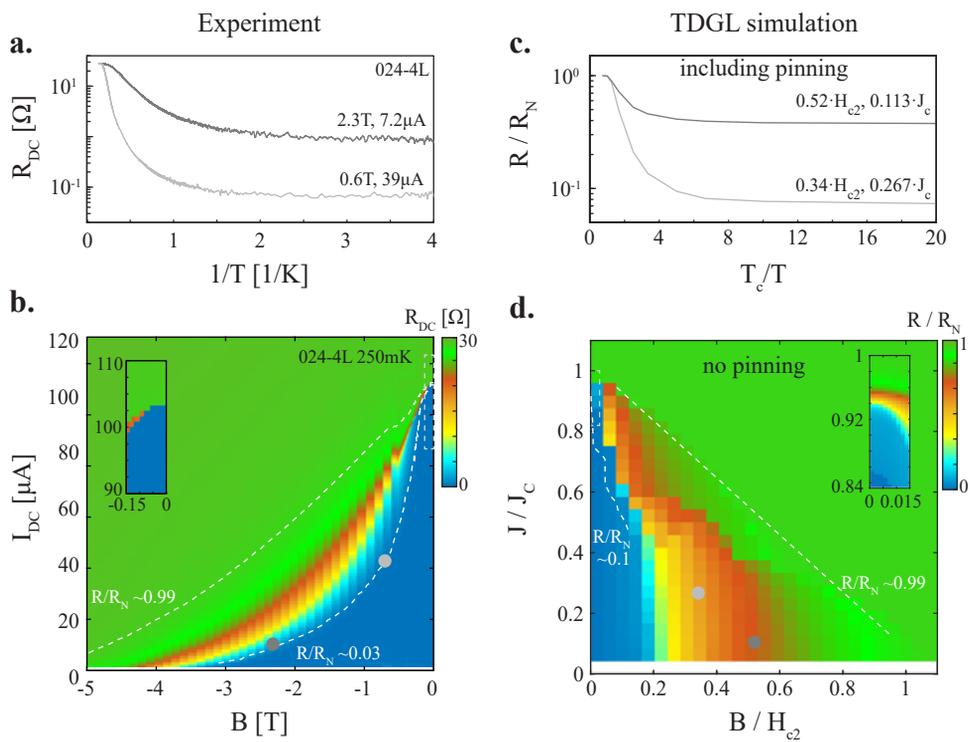

Figure 3

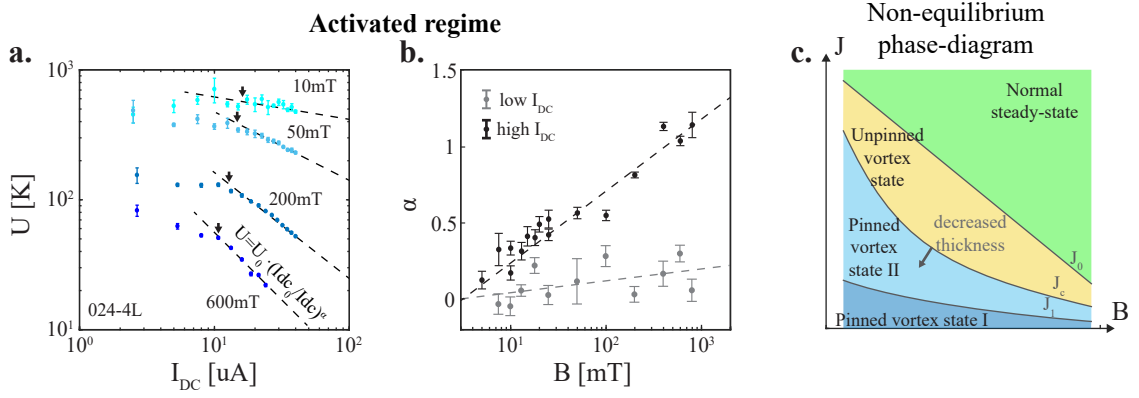

Figure 4

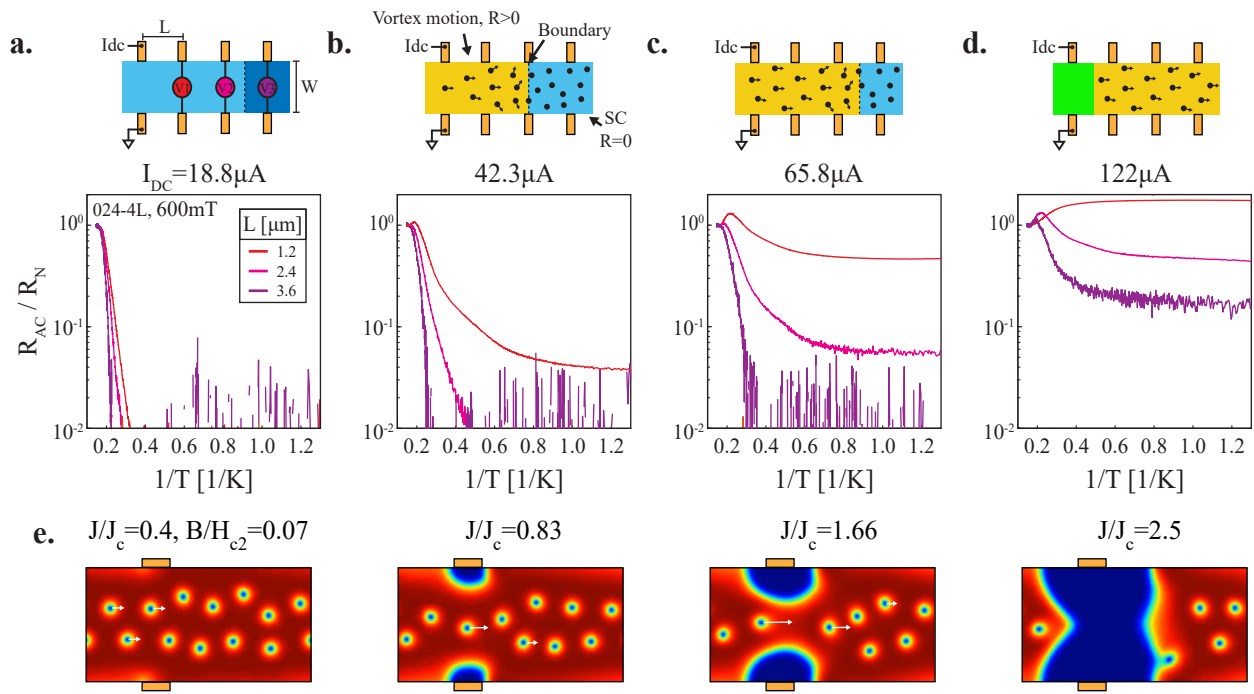

Figure 5

# Supplemental information for

# Absence of dissipationless transport in clean 2D superconductors

A. Benyamini[1†*], E.J. Telford[2*], D.M. Kennes[3], D. Wang[2], A. Williams[2], K. Watanabe[4], T. Taniguchi[4], J. Hone[1], C.R. Dean[2], A.J. Millis[2], A.N. Pasupathy[2]

## S1 - Device parameters

The table below summarizes the parameters for the three main devices shown in the paper. Each device had multiple contacts that showed similar results. For information on other measured devices see S3.

| Device | Layer # | $R [\Omega]$ | $Tc\ [K]$ | $H_{c2}\ [T]$ | $\xi\ [nm]$ | $l\ [nm]$ |
|---|---|---|---|---|---|---|
| 002 | 1 | 71 | 3.5 | 2.3 | 11.8 | 49.1 |
| 003 | 2 | 36 | 5 | 4.2 | 8.5 | 48.5 |
| 024 | 4 | 28 | 6.2 | 5.3 | 7.4 | 31.1 |

## S2 – Heating discussion

Joule heating of micron size 2D superconductors can happen due to finite resistance at finite temperature of the SC or due to finite contact resistance at the interface between the embedded gold electrodes and NbSe2. The measured 4-probe sheet resistance of few layered NbSe2 is $R_\blacksquare \sim 17 - 110\Omega$ depending on layer number and the contact resistances are of the order of $100's$ of ohms.

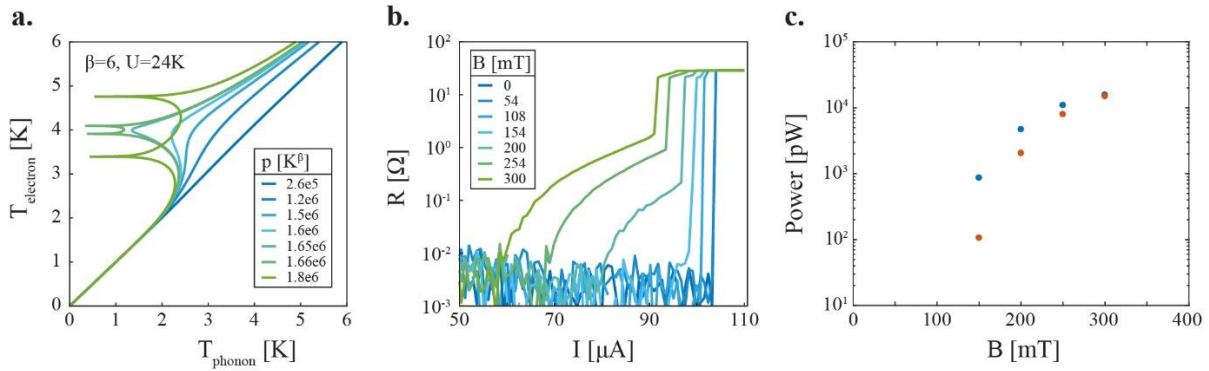

**Figure S1 – Heat balance equation and heating at jumps. a.** Solution of the heat balance equation for $\beta = 6$ and $U = 24K$ for varying $p \equiv R_0 I^2 / A$. For low $p$ $T_{el} \sim T_{ph}$ for all temperatures, as $p$ is increased a 'hump' is observed around a finite $T_{ph}$ which turns into an instability with two stable solutions. At low $T_{ph}$ the two stable solution are $T_{el} = T_{ph}$ and $T_{el} = T_{saturation} > T_{ph}$, showing that a heat balance equation can create saturation in the sample for high enough power. One feature we do not observe in experiment is the jump as a function of $p$ of $T_{el}$ to $T_{ph}$. **b.** Line traces of $R_{DC}$ versus $I_{DC}$ at low magnetic field at $T = 250mK$. Jumps are observed at a finite current. **c.** Summary of the power, $P = R_{DC} \cdot I_{DC}^2$, at the jump point for different magnetic fields. The blue and red dots correspond to the two sides of the observed hysteresis, see S3.

To calculate the heating of the SC a heat balance equations is needed[1],

$$P = A \cdot \left(T_{el}^{\beta} - T_{ph}^{\beta}\right),$$

where $P = R(T) \cdot I^2$ is the power coming in, $R(T) = R_0 \cdot \exp\left(-\frac{U}{k_B T}\right)$ is the resistance of the SC, $U$ is the activation energy, I is the sourced current, $A$ is a conversion factor, $T_{el}$ will be the temperature of the SC, $T_{ph}$ will be the temperature of the main source for thermal equilibration and $\beta$ is the exponent. This formalism will give saturation at low temperature for a finite power. The fits to our data sets work for $U \lesssim 10K$ which are measured only close to $H_{c2}$. Another feature of the heat-balance equation which isn't observed in the measured data is a jump in $T_{el}$ as the power gets to a critical value, see figure S1a at $T_{ph} \rightarrow 0$.

To check if the jumps observed for the 4-layer device are due to trivial heating we plot the power at the jump point for several magnetic fields, figure S1b-c. In panel b we show the resistance in log10 at high currents. In panel b we show the inferred power at the jumps. The power is evidently not constant at these points suggesting that a simple heating picture is not enough.

In the case of heating from the contact resistance, assuming again that thermalization happens through the contacts, we would anticipate that the sample would heat uniformly, a fact that we do not observe in the non-local measurement shown in the manuscript in figure 6. Another experimental evidence is that we do not see a change in the normal state resistance which is temperature dependent at higher temperatures above $T_c$.

Summarizing, although heating may be the main source for observed effect, the non-equilibrium phase diagram exhibits a wealth of physics which goes beyond a heat-balance equation. We have further results that are out of the scope of this paper and will be published independently that show a physical effect that cannot at this point be connected to heating. As our TDGL simulations, see main text and S3-4, exhibit most of the measured features we work under the assumption that vortex physics is the correct picture, albeit heating may still play a role at higher current, but not the dominant one.

## S3 – Activated behavior at equilibrium

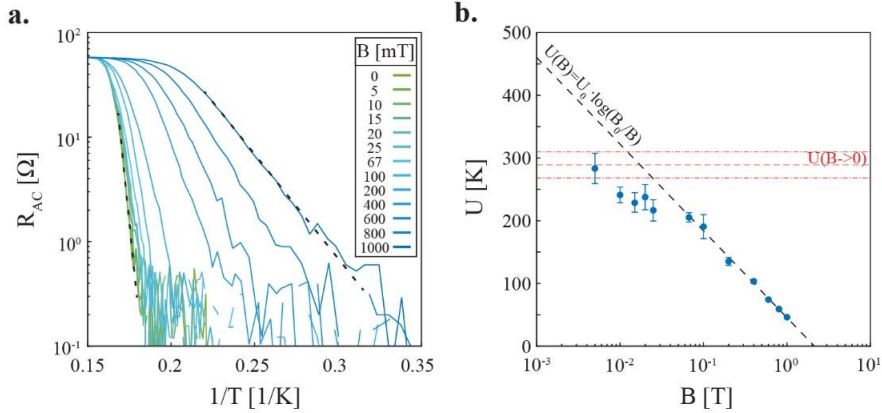

**Figure S2 – Activated behavior sensitive to magnetic field. a.** Temperature dependence of the resistance measured at equilibrium with a small AC current for different magnetic fields. A linear slope is seen for all traces in $\log R$ verus $1/T$ indicating activated behavior. **b.** Summary of the fitted activation energies versus magnetic field in log-scale. A cross-over is observed around $50 mT$ to a logarithmic dependence as expected from the long-range vortex interactions in 2D. The red dashed line is a guide for the eye of the activation extracted at $B = 0$, second dashed lines are the errorbar amplitude.

## S4 – Summary of critical currents

Figure S3a summarizes the critical current densities, $J_{c/0} = \frac{I_{c/0}}{Wt}$ ($W$ the flake width), dependence on layer number for all measured devices in log-scale at $B = 0$. Overall an exponential dependence is observed for the lower critical current density and a weaker dependence for the upper critical current density. Both critical currents converge at four layers. Theoretically $J_0$ can be due to the cooper pair breaking current density, $J^{pb} = \frac{n_e e \Delta}{m v_F}$ ($n_e$ is the superfluid electron density, $\Delta$ the SC gap, $m$ the mass of the carriers and $v_F$ the Fermi velocity), or due to the Ginzburg-Landau thin-film critical current density, $J^{GL} = \frac{H_{c2}}{6\sqrt{6}\pi\kappa\lambda}$. At $B = 0$ $J_0$ is roughly $\sim 10^{10} A/m^2$ two order of magnitude lower than an estimate of $J^{pb}$ but of the order of $J^{GL}$ which we associate to $J_0$. Figure S3b shows $J_0$ at $B = 0$ as a function of $H_{c2}$ with a linear fit to $J^{GL}$ giving $\lambda \sim 280 - 310 nm$ (for $\xi_\parallel = 8 - 10 nm$) on the order of $\lambda_{bulk} = 250 nm$.

At sufficiently large magnetic fields, a substantial number of vortices exist in the SC sample, which can be pinned either collectively or by disorder. We can attempt to understand dissipation at these fields in terms of the motion of these vortices. In this picture, the lower critical current density, $J_c$, is indicative of the minimal Lorenz force needed to free pinned vortices. At intermediate magnetic fields we observe that $J_c < J_0$ for all sample thicknesses studied here, as expected for a type-II SC with weak vortex pinning[2,3]. Two possible limits exist for the depinning force, depending on whether it is single vortices or a collectively pinned vortex bundle that is being depinned[2,3]. We can convert $J_c$ to these two limiting forces. In the limit of single vortex depinning, the critical force acting on a single vortex is $F_{SV} = J_c \phi_0 t$, while in the collective limit, the force acting on all vortices collectively is $F_N = N_V \cdot F_{SV}$ ($N_V = \frac{BA}{\Phi_0}$ is the number of vortices in the sample, $A = LW$ is sample area and $L$ is the sample length). The two conversions from Jc to force are shown in figure S3c as dashed and full lines respectively. The blue, red and black traces represent the data for the monolayer, bilayer and quadlayer device respectively. The correct depinning force will depend on the size of the vortex bundle which should increase at larger magnetic fields[2] and reside between the two limits, see illustrations in the figure. Measurement on other samples of corresponding layer number show similar force magnitudes, though the critical magnetic field varies between samples which shifts the position of the curves. We find an exponential increase of the vortex depinning force with increasing layer number. We postulate that the increase is due to enhancement of SC with layer number due to the tunnel coupling between layers or due to correlated pinning between layers as observed by the increase in $T_{c2}$, the SC gap or the superfluid stiffness with layer number.

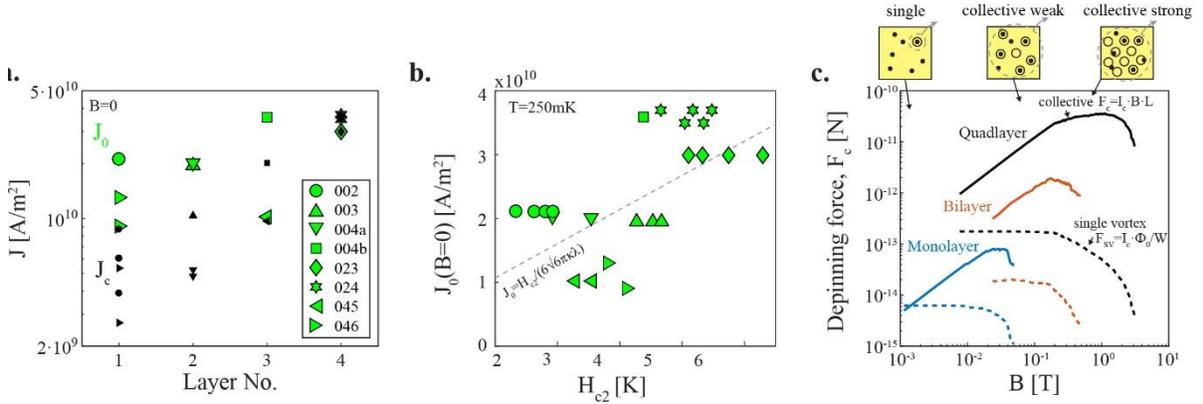

**Figure S3 – Behavior of the critical currents in the few layer limit. a.** Summary of the critical current densities for various devices plotted versus layer. **b.** Dependence of $J_0$ at $B = 0$ on $H_{c2}$. A linear fit to $J_0 = \frac{H_{c2}}{6\sqrt{6}\pi\kappa\lambda}$ is shown. **c.** Conversion of $I_c$ to the force acting on the vortices in the two limits where the force acts on a single vortex, dash lines, and on all vortices in the sample, full line. The conversion formulas are noted and illustrations for the vortex bundle state which will get depinned at the lowest current.

## S5 – TDGL simulations – No pinning

We use deterministic TDGL equations[4–6] to describe the dynamics of the 2D superconductor in the presence of both a magnetic field as well as a sourced current. The key quantities to monitor are the complex superconducting order parameter $|\Delta|e^{i\phi}$, the charge density $\rho$ as well as the current density $\vec{j}$. The electromagnetic fields are represented by the vector potential $\vec{A}(t)$ and scalar potential $\Theta(t)$. We choose units by $\hbar = c = e = 1$ which means that the superconducting flux quantum $\Phi_0 = \frac{hc}{2e} = \pi$.

The equations to solve are

$$\frac{1}{D}(\partial_t + 2i\Psi)\Delta = \frac{1}{\xi^2\beta}\Delta[r - \beta|\Delta|^2] + \left[\vec{\nabla} - 2i\vec{A}(t)\right]^2\Delta$$

$$\rho = \frac{\Psi - \Theta}{4\pi\lambda_{TF}^2}$$

$$j = \sigma\left(-\nabla\Psi - \partial_t A(t)\right) + \frac{\sigma}{\tau_s}Re\left[\Delta^*\left(\frac{\nabla}{i} - 2A\right)\Delta\right]$$

Here $D$ is the normal state diffusion constant, $\Psi$ is the electrochemical potential per electron charge, $\xi = \sqrt{6D\tau_s}$ is related to the superconducting coherence length $\xi_0 = \xi/\sqrt{\frac{r}{\beta}}$, where $\tau_s$ is the spin-flip scattering time, $\lambda_{TF}$ is the Thomas-Fermi static charge screening length and $\beta$ is a system dependent constant that sets the magnitude of the order parameter. For definiteness we measure lengths in units of $\xi$ and time in units of $\frac{\xi^2}{D}$ (which we write simply as $D^{-1}$ since $\xi$ is our unit of length) and choose parameters $\beta = 1$, $\sigma = 0.1$, $\tau_s D = \frac{1}{6}$ and $\frac{\lambda_{TF}^2}{\xi^2} = 0.1$. We chose those units for definiteness, but verified that none of the general conclusions is lacking. To close this set of equations we supplement them with the continuity equation

$$\partial_t \rho + \vec{\nabla} \cdot \vec{j} = 0$$

and the Poisson equation for the scalar potential

$$\nabla^2 \Theta = -4\pi\rho$$

Using a finite elements approach we solve the coupled partial differential equations with periodic boundary conditions in the y-direction while being open in the x-direction (choosing first derivatives to vanish at that boundary as a boundary condition). We discretized time in steps of $D\Delta t = 0.0001$, but verified numerically that the results obtained are converged upon decreasing this numerical parameter further. For finite sourced current we choose the boundary conditions of the current density such that the one end of the open boundary acts as a particle source while the other acts as a drain $j_x(x = 0, y, t) = j_x(x = L, y, t) = j_0$ with L the size of the system. The initial conditions for all other variables but the order parameter $\Delta$ are chosen to be zero at $t = 0$, while for $\Delta(x, y, t = 0)$ we choose initial values drawn from a random uniform distribution in the interval [0,0.001]. For given external magnetic field $B$ and zero external electric field we determine $A$ from $\nabla \times A = B$ assuming Coulomb gauge.

We concentrate on a geometry which has open boundaries along the *x* direction and periodic boundary conditions along the *y* direction. The system we consider is thus a torus. However, when vortices move in a slap geometry with open boundaries vortices are destroyed at the one end and created at the other giving similar physics.

**S6 – Simulating B-T phase diagram at equilibrium with no pinning**

We find that TDGL simulation of the B-T phase diagram, figure S4b (see also S5), with no pinning and no thermal fluctuations qualitatively reproduces the transition from the normal state to the activated region seen in the experiment, figure 1d and S4a. Due to the absence of pinning and thermal fluctuations in the TDGL simulation, the activated behavior is not captured well in this model. However, the dependence of the critical field on temperature is captured well. One of the findings of the TDGL simulation, see supplemental movies M1-M3, is that for samples of the sizes typically achieved in exfoliated monolayers, edge effects are of importance, and the details of the sample geometry play a significant role in the shape of the critical field line as a function of temperature.

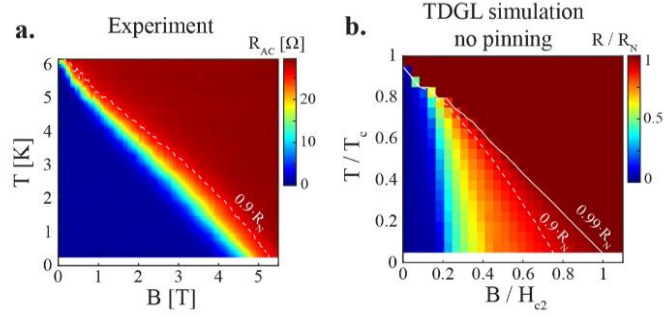

**Figure S4 – Comparison of TDGL simulation with no pinning to experiment. a.** Experimental phase diagram for the quadrilayer device as shown in the manuscript, figure 1. **b.** TDGL B-T phase diagram.

## S7 - Simulating TDGL including disorder

To include disorder we generalize the above equations, S4, by replacing $r \to r(x,y)$. We rewrite $r(x,y) = r_0 f(x,y)$ with $r_0$ setting the superfluid stiffness without disorder and $f(x,y)$ describing the disorder effects. As $f(x,y)$ we choose

$$f(x,y) = 1 - \sum_{i=1}^{N} \delta_i \cdot \exp\left(-\frac{(x-x_i)^2}{\zeta^2} - \frac{(y-y_i)^2}{\zeta^2}\right),$$

where $N$ describes the total number of defects, $(x_i, y_i)$ denotes the position of the $i$-th defect and $\delta_i$ is the $i$-th defect's strength. We draw $x_i$ and $y_i$ from a uniform distribution $(0, L]$ as well as $\delta_i$ from $(0, \delta_{max}]$.

## S8 – Hysteresis at low B

Hysteresis in measured resistance is observed with current sweeps at low fields. Figure S5 shows on the right the full measured diagram with a dark line showing the $0.01 \cdot R_N$ and $0.99 \cdot R_N$ resistance contours. The finite magnetic regime above $\sim 350 mT$ show no hysteresis. The two left contours show zoom-ins on lower and lower magnetic field regimes at higher absolute currents. This lower regime exhibits a jump in the resistance which is hysteretic. This is observed by the differences in the positive and negative direct currents scans and by the black line shown in the upper middle panel representing the position where the jump occurs in the lower middle panel.

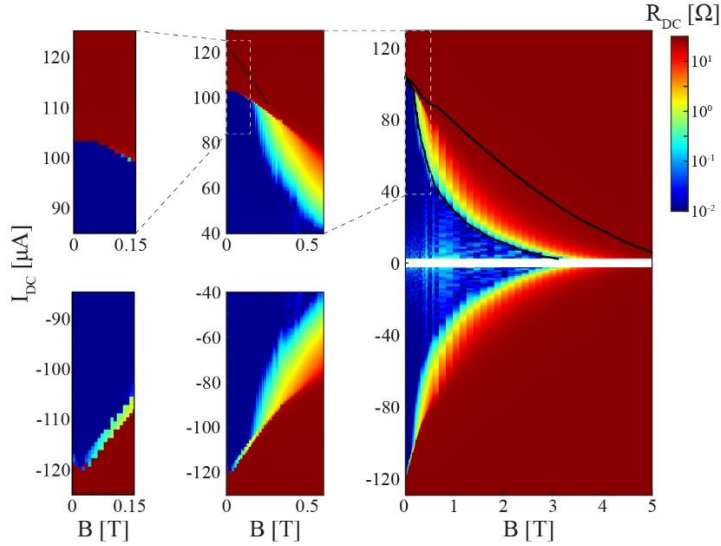

**Figure S5 – Hysteresis of resistance jumps.** Resistance, $R_{DC} = V_{DC}/I_{DC}$, colormap in log-scale. The right panel shows the full measurement done by sweeping positive to negative currents. The data around zero current is removed due to artifacts of division by zero.

## S9 – Temperature dependence at non-equilibrium

The non-equilibrium, finite DC current, behavior is shown in figure S6a-c for three magnetic fields representative of the different observed physical regimes. The blue to green traces show $R_{DC}(1/T)$ for the same current range on all plots. At $10mT$ the activated behavior only weakly depends on the current amplitude while at $600mT$ it depends strongly and stops behaving activated at intermediate currents. At higher currents at $200mT$ and $600mT$, yellow to red traces, shows saturation of $R_{DC}(1/T)$ as $1/T \to \infty$, while for $10mT$ no such behavior is observed. Jumps are observed both at $10mT$ and $200mT$ which reduce in amplitude and disappear at increased magnetic field.

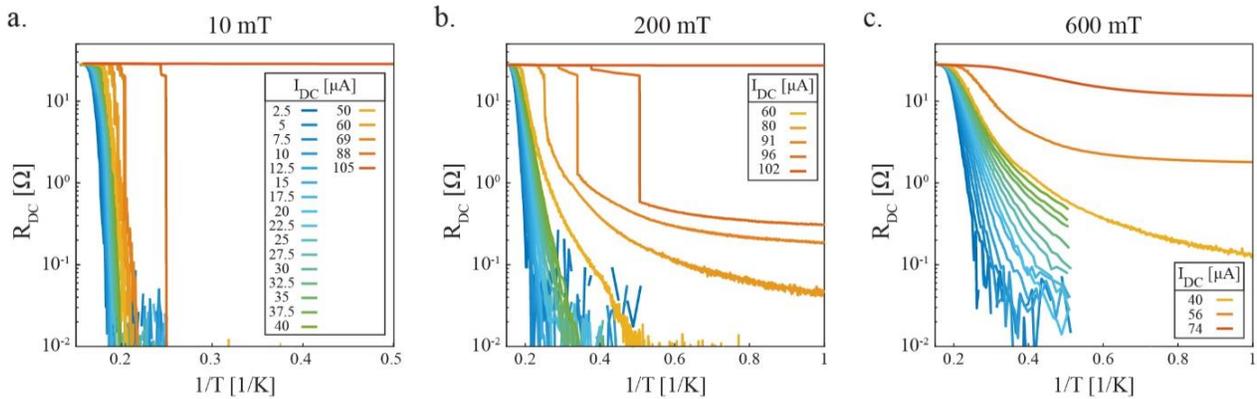

**Figure S6 – Temperature dependence of resistance at non-equilibrium. a-c.** $R_{DC}$ shown in log-scale vs inverse $T$ for varying $I_{DC}$ for three representative magnetic fields $10mT$, $200mT$ and $600mT$. Blue to green traces represent the same $I_{DC}$ in a-c. Yellow to red represent larger currents as noted in each legend.

Figure S7 summarizes phenomenologically the observed phases at our lowest measuring temperature, $T\sim 250mK$. The dashed lines are inferred from resistance contours from figure S5, the colored circles are from the different observed crossovers shown in figure S6 between the two activated regimes and between the activated to saturated regimes. The grey area is shown for currents we cannot associate with activation or saturation behaviors.

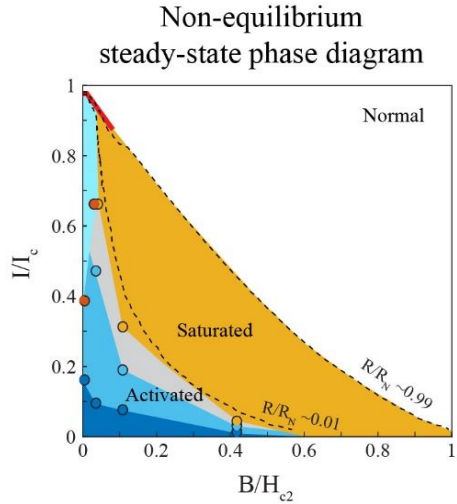

**Figure S7 – Non-equilibrium phase diagram.** Dashed lines and colored circles are inferred from measurements. The underlying colors represent the different phenomenological regimes discussed in more detail in the supplementary and the manuscript.

To clarify the temperature dependence, we draw three phase diagrams showing the inferred physical regimes as a function of temperature and current for three different magnetic fields. The phase diagrams are shown in figure S8. The cross over from the normal state is shown by the black contour in case of a continuous change in resistance, while for a jump in the resistance a red line is shown (for hysteresis see S5). The point which we get to the noise floor is shown by the black dashed line. A further red dash line is shown to shown when a discontinuity is observed below the continuous drop form the normal state. For the lowest magnetic fields, left phase diagram, we cannot extrapolate what is the nature of the physical state below the jump and it is shown in turquoise.

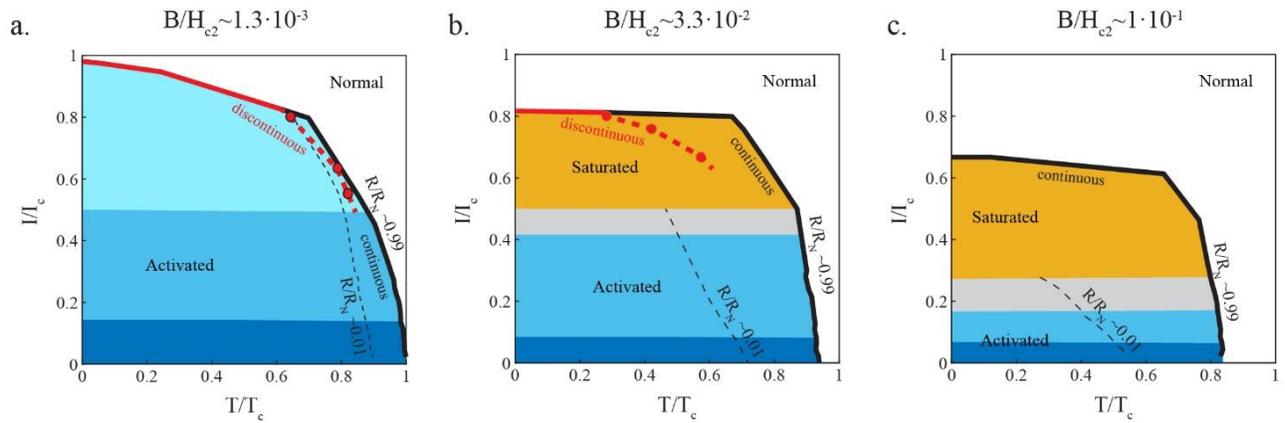

**Figure S8 – Non-equilibrium I-T phase diagrams. a-c.** Phase diagrams summarizing S6 and S7.